\documentstyle[12pt,psfig]{article}
\begin{document}
\topmargin = -1.0 cm
\leftmargin = 1.0 cm
\baselineskip .3in
\sloppy
\newpage
\pagestyle{plain}
\def\be{\begin{equation}}
\def\ee{\end{equation}}

\title {Aspects of Dielectric Breakdown in a Model
for Disordered Nonlinear Composites}

\author{Abhijit Kar Gupta and Asok K. Sen$^*$ \\
{\it LTP Section, Saha Institute of Nuclear Physics}\\
{\it 1/AF Bidhannagar, Calcutta 700 064, India}}

\date{\today}
\maketitle

\begin{abstract}
We study dielectric breakdown in a semi-classical bond percolation model
for nonlinear composite materials introduced by us and the related
breakdown exponent near the percolation threshold in two dimensions.
The breakdown exponent after doing finite size scaling analysis is 
found to be $t_B \simeq$ 1.42.  We discuss in detail the differences in
our model from the traditional models for dielectric breakdown and argue
that our result seems to be different from the standard result of 4/3
obtained in the previous models.

\end{abstract}


$^*${\it Corresponding author, e-mail address: asok@hp2.saha.ernet.in}

\newpage

\section*{Introduction}

\vskip 0.2 in

Statistical physics of the breakdown of an insulating dielectric into 
a conducting state (or of a conductor into a `fused' insulating state) has been
the subject of intense research \cite{frac} for more than a decade now.
Suppose one considers a random binary (two-phase) mixture of metallic and 
non-metallic components. 
If the volume 
fraction of the metallic phase is large enough, the metal phase forms 
at least one \cite{mini} sample-spanning cluster in which the non-metallic 
phase is dispersed in the form of isolated islands. In this regime the 
electrical conductivity of the sample is large. The system is {\it metallic}. 
On the other hand, for a small 
volume fraction of the metallic components, the non-metallic phase forms 
at least one \cite{mini} sample-spanning cluster in the presence of small
and isolated metallic islands. 
The system is then in the {\it dielectric} or insulating regime. The electrical 
conductivity of the sample in this regime is ideally zero and extremely small 
in practice. Now, if one increases the electric field across the sample in this
regime, the voltage across the non-metallic bonds keeps increasing and it is
not unlikely that some of them may give in to let some current through them or 
turn metallic. Clearly in this case the breakdown problem is set up with an
underlying percolation model.

In the usual dielectric breakdown model \cite{bow} of a random mixture of
conductors and insulators it is assumed that each insulating bond can withstand
a fixed potential difference across it and becomes a conductor if the local
potential difference exceeds its threshold. Therefore, the whole lattice is subjected 
to breakdown at any volume fraction ($p < p_c$) of conducting 
components when an appropriately large external voltage, called the 
{\it breakdown voltage}, $V_B$, is applied.  Its value depends on the specific
configuration of the sample, and usually one talks about the 
configuration-averaged value of $V_B$ at any particular $p$.
For $p=0$, {\em i.e.}, when all the bonds are insulators, the breakdown 
voltage ($V_B$) scales as the linear size ($L$) of the lattice: $V_B/L = v_g$,
where $v_g$ is the voltage threshold (all insulators are assumed to have an 
identical voltage threshold) for an individual tunnelling bond.
For $p \geq p_c$, such a lattice is conducting for any small applied voltage
and the question of dielectric breakdown does not apply: $V_B/L = 0$. To remove
the trivial system size ($L$) dependence, we talk about the external breakdown 
field ($E_B = V_B/L$) instead of the breakdown voltage from now on. The
interesting thing happens as one approaches $p \rightarrow p_c$ from below.
One obtains a criticality and a power law
\be
E_B \sim (p_c - p)^{t_B},  \label{expo}
\ee
where $t_B$ is called the breakdown exponent. 
A similar scaling is also known for the mechanical fracture process \cite{ray}:
$\sigma_{min} \sim (p - p_c)^b$, where $\sigma_{min}$ is the minimum stress
needed to break the system apart. 

In general in a breakdown
process one defines two critical voltages: one is the breakdown initiation
voltage $V_I$ at which the nucleation of the breakdown process (akin to an
avalanche) is initiated and the other is the final breakdown voltage $V_F$
which is the minimum voltage at which the system as a whole breaks apart. 
In some cases $V_I$ is no different from $V_F$ \cite{bow}. In other cases
the system needs some more voltage beyond $V_I$ to reach the final breakdown state.
It is commented in some earlier works \cite{man, beale} that $V_I$ and 
$V_F$ are essentially the same. So the authors of many previous works had 
actually treated the average value of $V_I$ as the average breakdown voltage 
($V_B$). The statistics of $V_I$ and $V_F$ are also claimed to be the same 
{\em i.e.}, they are described by the same critical exponent near the 
threshold ($p_c$). There has been a number of works (see {\em e.g.} the 
references \cite{man, beale}) in the 
literature for estimating the {\it breakdown exponent}. 
A closely related quantity of interest is the {\it mimium-gap path}, $g(p)$, 
of a non-percolating lattice configuration.  It is the 
minimum number of insulating bonds which are to overcome their thresholds 
to give a connecting path between two opposite sides of a lattice across 
which the external voltage is applied. 
The breakdown voltage ($V_B$) and the minimum gap ($g$) are actually 
two different quantities \cite{man} except at $p$ = 0 and  at $p = p_c$ 
although both the quantities near $p_c$ seem to behave in the same way 
and the numerical
results \cite{man} claim that their scaling exponents are the same near 
$p_c$. It was claimed through an analytical calculation on a hierarchical
lattice and through a numerical simulation on a square lattice \cite{stin}
that the breakdown voltage ($V_B(p)$) behaves like $g(p)$ in a random lattice. 
The average of $g(p)$ is supposed to vary as 
$(p_c - p)^{t_g}$, where $t_g$ (the minimum-gap exponent) is identified with 
the breakdown exponent $t_B$. Later, it was rigorously established by Chayes 
{\em et al.} \cite{chayes} in an invasion (or, forced under a pressure)
percolation type situation, that $t_g = \nu$ in 2D, where $\nu$ is
correlation length exponent.
This indicates that $t_B = t_g = \nu$. However, there is a $logL$ 
term involved in the scaling relationship of breakdown field ($E_B$) near
percolation threshold ($p_c$), and $E_B \sim {(p_c-p)^{\zeta} \over lnL}$ 
\cite{beale}.

\vskip 0.2 in
\section*{The Model} 

\vskip 0.2 in

We propose a semi-classical (or, semi-quantum) model of percolation \cite{skg}
which works on the borderline between a classical and a quantum picture. 
Quantum physics enters our discussion through the possibility of tunnelling
of a charge carrier through a barrier (which does not exist classically).
Disorder in such systems is known to give rise to `pinning' or inhibition
to transport (say, in charge-density-wave (CDW) systems or flux-vortex 
lattices of type-II superconductors) upto a critical value of the applied
field above which tunnelling is active.
Our approach would be to solve an appropriate electrical network based on 
a semi-classical percolation model.  
Made of both random resistive and tunnelling elements, this network will be
called a random resistor cum tunnelling-bond network (RRTN).  There exists
a similar model in the literature called a dynamic random resistor network
(DRRN) proposed by Gefen {\em et al.} \cite{gef} to explain the crossover 
exponent in the experiment on $Au$-films reported in the same reference. 
The difference between these two models lies in the fact that the tunnelling 
elements (or the {\it imperfect} insulators) in the DRRN could be anywhere in
the non-metallic domain of the system whereas in the RRTN, these elements exist 
only in the proximity gap between two metallic domains (one can imagine that
the charge transfer by tunnelling should be most effective only in such gaps). 
Now tunnelling may take place through the
tunnelling bonds in various ways, so that the functional form of the
tunnelling current as a {\it nonlinear} function of the potential difference
across them may be quite complicated. For simplicity, we address
the aspects of nonlinearity in a macroscopic system which comes through
two piecewise linear regions of a tunnelling element.
The piecewise linear transport is in fact
a highly nonlinear process as there is a cusp singularity at the
intersection point. The transport due to tunnelling which is the source of
nonlinearity in the experimental systems \cite{rkc, chen}
we focus on, can be 
well approximated in this way and thus the nonlinearity of the
macroscopic systems may be understood at a qualitative (and, sometimes
even at a quantitative) level. Next, one notes as discussed above that in many
physical systems, the response is
negligibly low (or there is no response at all) until and unless the 
driving force exceeds a certain threshold value.
So a class of problems exist where sharp thresholds to transport occur. The
examples in the electrical case are a Zener diode, a CDW system or a typeII 
superconductor and in the fluid permeability problems, for example,
is a Bingham fluid (where there is a critical
shear stress $\tau_c$, above which it has a finite viscosity and below
which it is so enormously viscous that it does not flow).
In our RRTN model, we work with tunnelling
bonds which have zero conductance below a threshold.

But, our percolative model is not just a random mixture of two phases.  For
our convenience we take a square lattice in 2D. The basic physics should
remain the same if we go over to 3D.  Conducting/ ohmic bonds ($o$-bonds) are
thrown at random at a certain volume fraction $p$.  The rest $(1-p)$ fraction
contains insulators.  Now we allow tunnelling bonds ($t$-bonds) only  across
the nearest-neighbour ($nn$) gaps of two conducting bonds (and no
further) if an appropriate voltage is
applied externally across two opposite sides of such a random resistor cum
tunnelling network (RRTN). Our interest is to examine this proposed 
correlated percolation model in the spirit of dielectric breakdown phenomenon. 
The mechanism operating here is clearly not traditional dielectric breakdown 
because the piecewise linear response considered here in the tunnelling bonds 
is {\it reversible} in the sense that if the local voltage difference is 
lowered below the threshold a tunnelling bond becomes insulating again. This is 
an important point because if we would assume the process to be 
irreversible, then the irreversible conversion of one insulating element to a 
conducting one may trigger an {\it avalanche} effect.
Since a local current redistribution takes place in the reversible RRTN 
model as well whenever a dielectric turns metallic, avalanche may take place
in this model as well, but the avalanches may be more restricted in the RRTN 
than the traditional reversible models.
Similarly in the random fuse network \cite{arcan} one has the
{\it irreversibility} with respect to conductor $\rightarrow$ insulator
transtion with the increase of applied field.  Breaking (fusing) of one bond
in a certain path permanently (because of too much stress) may lead to an
increase of current density in the other paths and thus it may trigger an
{\it avalanche} effect.  Since a local current re-distribution takes place
in the reversible RRTN model as well whenever a dielectric turns metallic,
avalanche may take place in this model too, but the avalanches may be more
restricted in the RRTN than in the traditional `irreversible' models. 
In practice the `reversibility' situation is achieved when the charge transport
by tunnelling gives the most important contribution to the breakdown process 
than the permanent breakdown of the microscopic conductors/ insulators
inside the system.  One example of this is the experiment on dielectric
breakdown demonstrated by Benguigui {\em et al.} \cite{ben} where the authors 
consider a network of tunnelling diodes.  There are many other real systems
demonstrating this reversibility, {\em e.g.}, carbon-wax mixture \cite{rkc},
$Ag$-$KCl$ composite \cite{chen} and many other nonlinear composites where 
the macroscopic $I$-$V$ characteristic is reversible (no appreciable 
hysteresis effect).

Next we comment on the procedure for obtaining the breakdown voltage ($V_B$)
for the usual dielectric breakdown problem as understood from the references
above. The usual procedure to obtain the electrostatic voltage distribution 
at the nodes of
the networks in the non-percolating situation is to solve for the Laplace's 
equation ($\nabla^2 V = 0$). This procedure, when discretised on a square 
lattice and in the situation where the dielectric constant for all the bonds 
are assumed to be the same (pure 
dielectric), reduces to $v_0 = \sum v_i/4$, where $v_0$ is the voltage at any 
node and $v_i$'s are the voltages at the four nearest-neighbour ($nn$) nodes on
a square lattice. In our case, we approach the breakdown point 
from the conducting side and apply Kirchhoff's law for our problem which 
takes the form: $v_0 = \sum v_ig_i/\sum g_i$, where $g_i$'s are the 
conductances of the $nn$ bonds. Clearly this may be reduced to the discrete
Laplace's equation above had the $g_i$'s for all the bonds been 
essentially the same. 
Further, in the usual models, as soon as the voltage difference ($v_i - v_0$)
across an insulating bond exceeds its threshold value $v_g$, this bond is
turned into a `perfect' conductor for all later time (iterations) to come,
and $v_i$ is made equal to $v_0$.  On the other hand, in our model, even when
a $t$-bond has been broken (turned metallic), neither does it become a perfect
conductor nor does it carry any carry any current until and unless the
voltage difference across it exceeds $v_g$.  We believe that this is a
crucial difference and should be more akin to reality. 

As per our model is concerned we assume that the tunnelling bonds (the bonds
which break) may be placed only in the $nn$ gaps of two conducting bonds and
nowhere else.  It will be noted that because of the reversible nature of our
$t$-bonds and their finite thresholds $v_g$, rarely would $V_I$ be equal to
$V_F$ in our model.  Indeed we {\it do not} work with $V_I$ and actually identify $V_B$ as the average of the final
breakdown voltages $V_F$.  Hence a typical breakdown path in the RRTN model
consists of an actual number of the so-called `broken' bonds and does not
quite correspond to the minimum gap path except when $p$ is very close to
$p_c$.  If there are $n$ number of active tunnelling (or broken) bonds in the
minimum gap path having a threshold voltage $v_g$ for each of them, the
overall breakdown voltage $V_B = nv_g$.  It may be noted that this is also
the case with the dielectric breakdown experiment by Benguigui {\em et al.} 
\cite{ben} on an artificially constructed electrical 
network of resistors and light emitting diodes (LED). 
The initial breakdown voltage $V_I$ (at which at least one tunnelling bond 
breaks) is just $v_g$.  Very rarely (except for $p$ near $p_c$) one has $n=$1,
and $V_B = V_I$ in our model or in the above mentioned experiment by Benguigui 
{\em et al.}  As a demonstration,
we show here a typical configuration (see {\bf fig.~1}) of the lattice of
size $L$=10 at a volume fraction $p$=0.30 where just one breakdown path has
been formed. Indicated by the dashed lines are the number of broken $t$-bonds.
The path is explicitly seen to {\it not} be the minimum gap path.

One may notice another difference of our model from the usual models of  
dielectric breakdown problems so far studied (where the dielectric bonds 
can break at any place in the network) from the above demonstration.
There may be a series of broken bonds at more than $nn$ gaps of two
conducting bonds in the breakdown path in the usual model
(see, {\em e.g.}, the figures in ref.  \cite{beale}) but not in ours.
It is worth commenting here that the breakdown paths
generated by Benguigui {\em et al.} \cite{ben} are more akin to our model than 
the usual model.  This is because even though many more than one LED's seem
to be broken in series, in practice two consecutive LED's are connected by
metallic wires and hence do not correspond to breakdown over two or more
near neighbour distances (or lattice constants).  
The breakdown exponent ($t_B$) in this experiment was reported to be $\cong
1.1$, which is smaller than what is actually expected ({\em i.e.}, 4/3, the
exact value of $\nu$ in 2D). The difference may be attributed to the finite
size effect since a system of size 20$\times$20 was considered.

\vskip 0.3 in

\section*{Calculation of the Breakdown exponent in the RRTN} 

\vskip 0.2 in

Here we examine the dielectric breakdown phenomenon in our model as the onset 
of nonlinear conduction against applied field for $p \le p_c$.
Below the percolation threshold ($p_c$) there exists a number of metallic 
clusters, isolated from each other, but closely spaced.  The conductivity is a
sensitive function of applied electric field/ voltage \cite{skg} as new
conducting paths are created when the applied external electric 
field increases above the dielectric breakdown field ($E_B = V_B/L$) of the 
insulator. Note that the model has a percolation threshold at $p_{ct} 
\cong 0.18$ \cite{kgs} if all the tunnelling bonds overcome their voltage 
thresholds at 
the appropriate positions. So below $p_{ct}$ there is no sample-spanning
cluster of combined $o$- and $t$-bonds, and hence there is no conduction (on
an average) at any finite electric field according to the criterion set for
our model.

Thus three types of {\it configurations} arise in the regime $p_{ct} < p < p_c$:
\begin{itemize}
\item Some configurations which are already percolating with the
ohmic bonds only: they have {\it zero} voltage threshold macroscopically,

\item Some configurations which donot percolate with the ohmic bonds only but 
do so in conjunction with the tunnelling bonds: they have a {\it finite} 
voltage threshold,

\item Some other configurations are there which never percolate even with the 
assistance of all the available tunnelling bonds: they {\it do not take part} 
in the breakdown process.
\end{itemize}
This third possibility does not arise in the usual class of breakdown problems
where any insulating bond may break given enough voltage and hence eventually 
renders the system conducting. 

Clearly, to find the average breakdown voltage ($V_B$) we have to disregard
those configurations which do not take part in the breakdown phenomenon.
In the {\bf fig.~2}, a typical distribution of breakdown voltages $V_B$
is shown for a system size 
$L = $40 and $p = $0.45. This distribution is quite broad and seems to be
asymmetric. 

The phase diagram is shown in the {\bf fig.~3} as the average of 
breakdown voltage ($V_B$) plotted against the volume fraction ($p$) of 
conducting bonds. This typical 
figure is shown for a system of size $L$ = 30 and average is taken over 
500 configurations. Our interest would be to know how does the average breakdown
field ($E_B = V_B/L$) scale against $(p_c - p)$ as in eqn.~(\ref{expo}). 
One usually plots the quantity $V_B$ or $E_B$ for a finite sized system 
against $(p_c - p)$ around $p_c$ in
log-log scale and find out the breakdown exponent $t_B(L)$ from the least
square fit. To remove the finite-size effects, however, we follow a slightly
different way of finding the above exponent. We first obtain the finite size
scaling of the breakdown field, $E_B$. One such scaling plot is shown in
{\bf fig.~4} for $p=$0.4.  In this way,
we obtain the asymptotic values $E_B(L=\infty)$ of the breakdown
field for all $p$ ranging from 0.3 to 0.5 through finite size scaling, 
which seems to follow
\be
E_B^p(L) = E_B^p(\infty) + a(p)L^{-\mu(p)},   \label{ebfsize}
\ee
where $\mu(p_c) \simeq 1$; but quite different (0.4 to 0.75) at other $p < p_c$.
Further $E_B^{p_c}(\infty)$ has a very small but positive value which
for the accuracy of our calculation implies that $E_B^{p_c}(\infty) = 0$. 
But as $p$ becomes smaller and smaller than $p_c$, $E_B^p(\infty)$ increases
systematically as the graph in {\bf fig.~5} indicates. We point out that 
forcing $E_B^p(\infty)$ = 0 at $p < p_c$ gives significantly worse fitting.
Eqn.~(\ref{ebfsize}) strongly demonstrates the fact that the breakdown model
we are considering is somewhat different in nature from the usual models
available in the literature where one observes a $1/lnL$ scaling of $E_B^p$
demonstrated clearly in the work of Beale and Duxbury \cite{beale}. This
scaling, which makes the $E_B^p$'s vanish irrespective of the $p$ in a truly
infinite size system, is non-existent in our model.  Since the breakdown
field in the previous models vanishes to zero irrespective of any $p$
($p<p_c$), it is worth noting that the above $1/lnL$ scaling and the
consequent vanishing of $E_B^p$ is also non-existent in another model which
has no dilution but has reversible tunneling conductors with random thresholds
at each and every bond in the lattice.  In such a network, Roux and Herrmann
\cite{rh} found that $V_B=(0.22 \pm 0.02)L$.

The scaling of the asymptotic breakdown field $E_B^p(\infty)$ can be 
written as 
\be
E_B^p(\infty) \sim (p_c - p)^{t_B}. \label{expo1}
\ee
The double logarithmic plot of eqn.~(\ref{expo1}) is shown in {\bf fig.~6}
and the least square fit of the data is also shown.  We find from this
fitting that the breakdown exponent $t_B \cong 1.42$ for our RRTN model.

\vskip 0.3 in

\section*{Discussion} 

\vskip 0.2 in

It seems that the above exponent $t_B$ is not very different from that of the
usual breakdown exponent $t_B=\nu=$1.33 as discussed above.  But it is not
unlikely either that we do indeed have a different result in our hands.
If different, it could be because of the nature of the electric field in
increasing the effective volume fraction of the conductors.  As may be
understood, the electric field adds on new bridge bonds (active $t$-bonds)
at well-determined positions (according to the {\it deterministic} laws of
electricity) which must be completely different from the {\it random} positions
of the extra $o$-bonds obtained by increasing the volume fraction to the same
effective volume fraction as obtained by applying the electric field.
Intuitively, the correlations obtained by these two different means
should be qualitatively different (one being isotropic and the other
anisotropic).  Indeed, as seen in an experiment \cite{ben} as well as in
simulations \cite{beale, kgs1} (see also {\bf fig.~1}), an electric
field tends to make somewhat elongated clusters directed towards the direction
of the external field.  But, our results in ref.~\cite{kgs1} do also show
that while the directivity (anisotropy) of the clusters increases with an
increasing field upto a maximum, it does finally start to decay (i.e., grows
more and more isotropic) at still larger fields and the RRTN at an infinite
field, which becomes our fully correlated bond percolation model \cite{kgs},
does not fall in the category of directed percolation (rather it falls in the
same universality class as the ordinary random bond percolation). Thus, at
a small but finite field, we may observe the percolation statistics to be
directed only a little bit.  Now, it is well-known from the results on
directed percolation that the correlation length exponent in a direction
parallel to the electric field is $\nu_{\parallel} \cong$ 1.7. So, it is
{\it not unlikely} that the correlation length exponent near the breakdown
field (which is quite small) takes some value between 1.3 and 1.7.  If true,
this may very well explain why out $t_B$=1.42.  In this respect, it may be
noted that Beale and Duxbury \cite{beale} also found the average $t_B=$1.46.
Thus, the exponent $t_B=$1.42 for our model may actually be a result
different from the standard quoted result of 4/3 for this exponent.  As a
final remark we would like to add that
it would be worthwhile to take the usual model for breakdown and repeat the
calculation for the final breakdown exponent under the condition that the
dielectrics are reversible ({\em i.e.}, that they are not broken permanently)
and compare the breakdown exponent with the one obtained here.

\newpage
\centerline{\bf Figure Captions}

{\bf Fig.1}
A typical configuration of the lattice for a square of size $10\times10$ 
with $p$ = 0.3 (below $p_c$). The breakdown path is indicated
by `$abcd$' with $n = 4$ which is seen to be different from the minimum-gap 
path `$aef$' of an usual dielectric breakdown model with $g(p)$ = 3.

{\bf Fig.2}
A typical distribution of breakdown voltage $V_B$ with $v_g$ = 0.5.

{\bf Fig.3}
The phase diagram for the dielectric breakdown is shown with the average of 
breakdown voltage ($V_B$) plotted against the volume fraction ($p$) of 
conducting bonds for a square lattice with $L$ = 30.

{\bf Fig.4}
The finite size scaling of breakdown field $E_B$.
 
{\bf Fig.5}
The behaviour of asymptotic breakdown field $E_B^p(\infty)$ with $p$.

{\bf Fig.6}
The log-log plot of $E_B^p(\infty)$ against $(p_c - p)$ and the best fit 
line to find the breakdown exponent. The fitted line gives $t_B$ = 1.42.
\newpage
%
%
\begin{figure}
\psfig{file=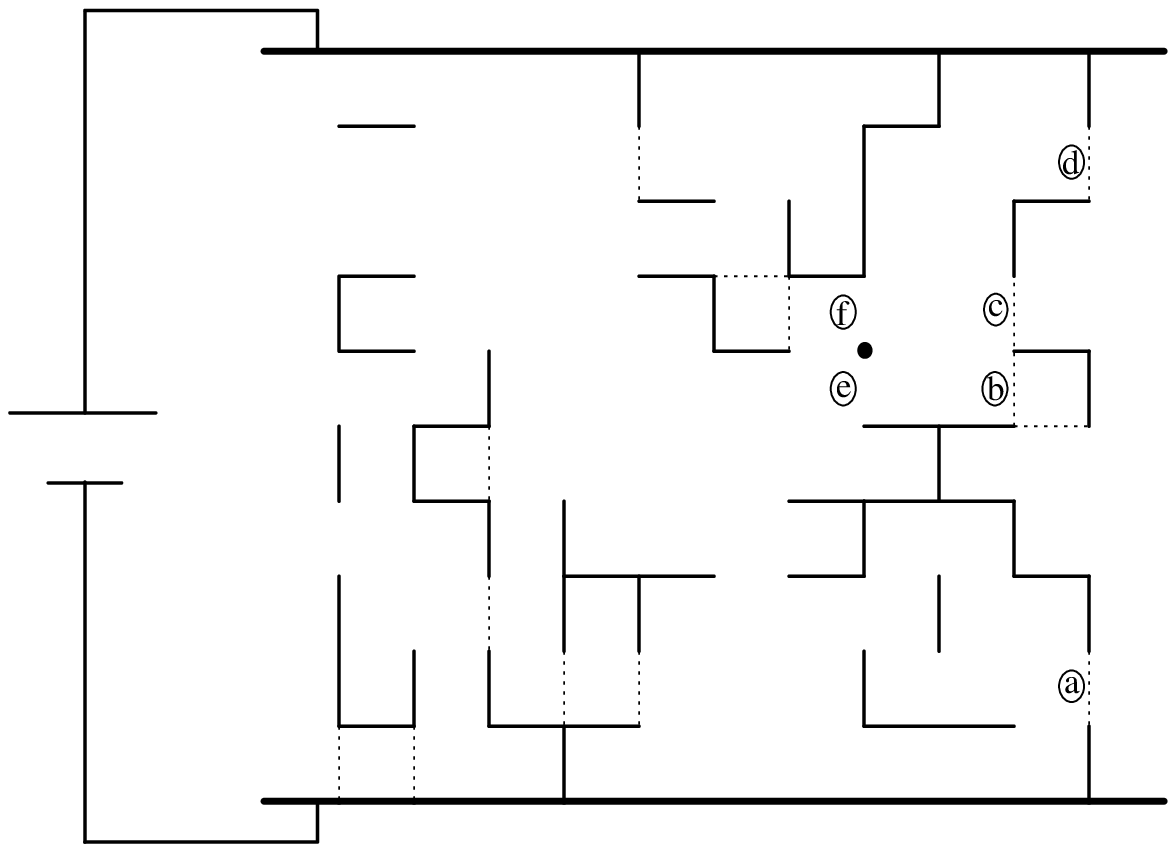}
\end{figure}

\begin{figure}
\psfig{file=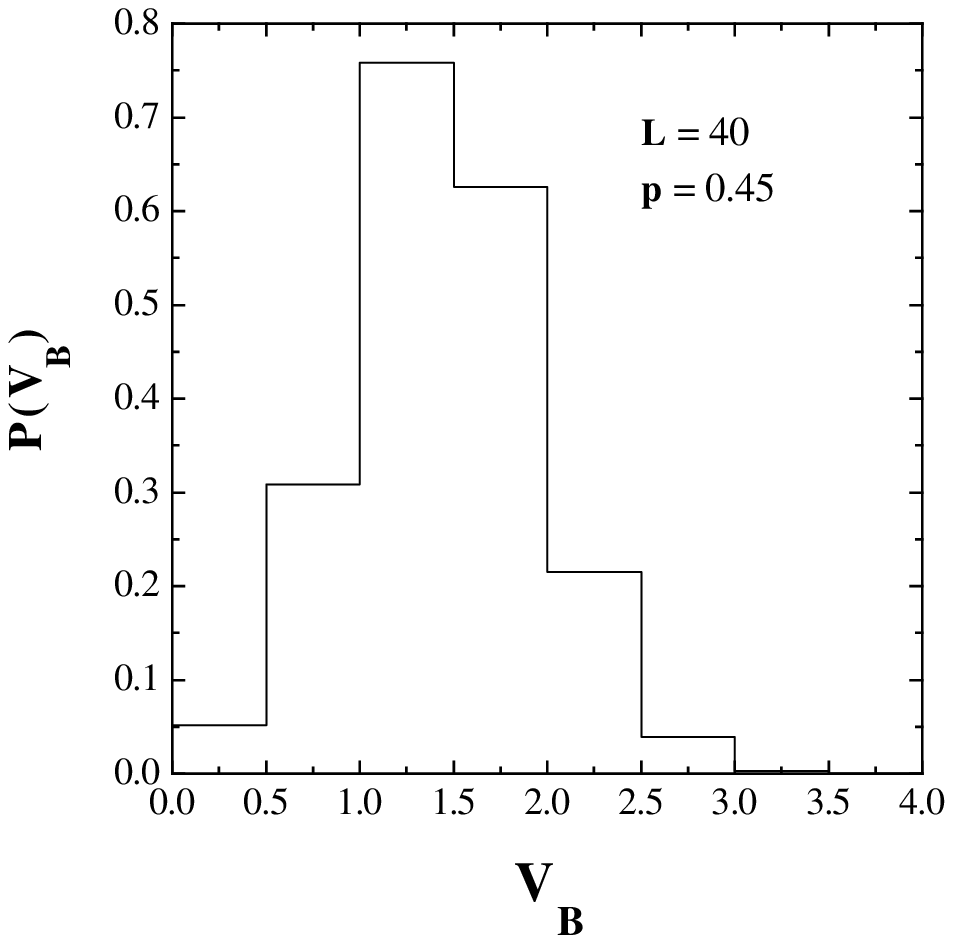}
\end{figure}

\begin{figure}
\psfig{file=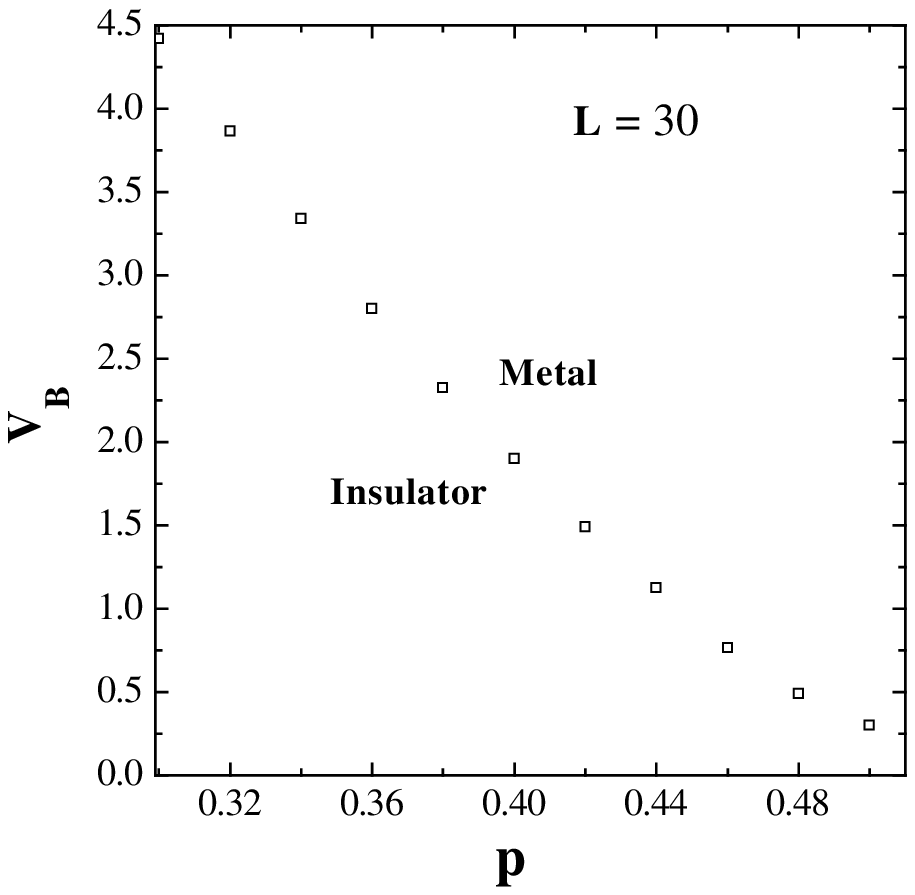}
\end{figure}

\begin{figure}
\psfig{file=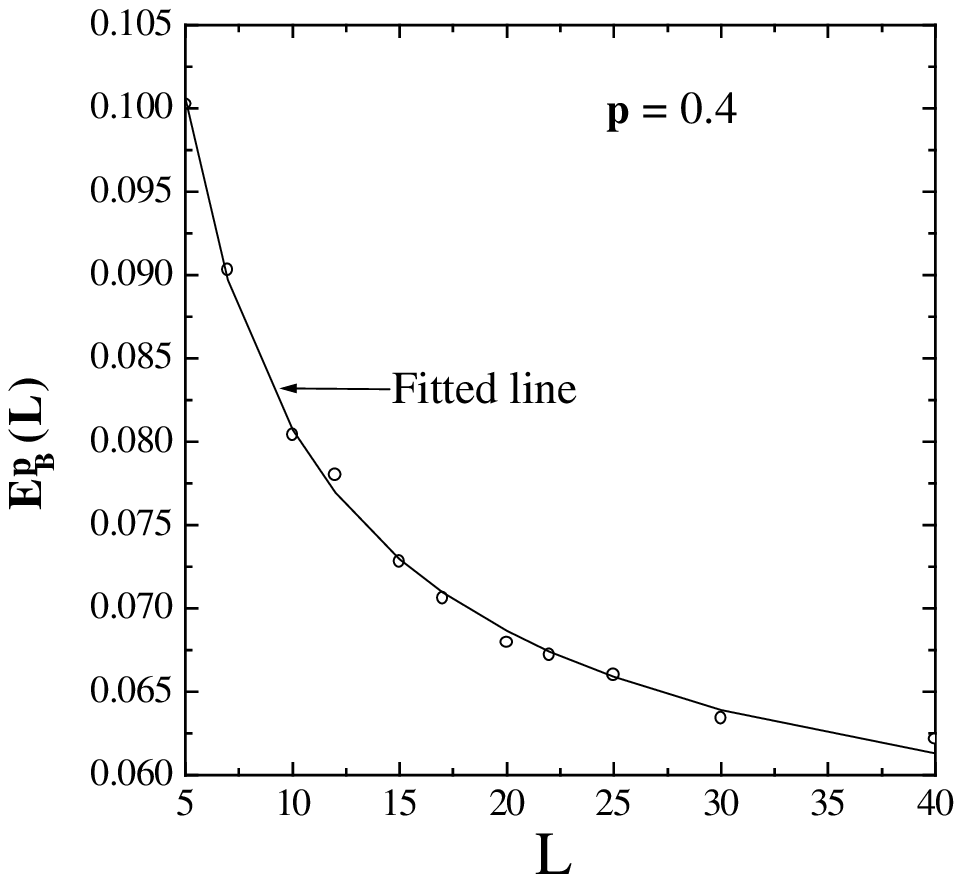}
\end{figure}
 
\begin{figure}
\psfig{file=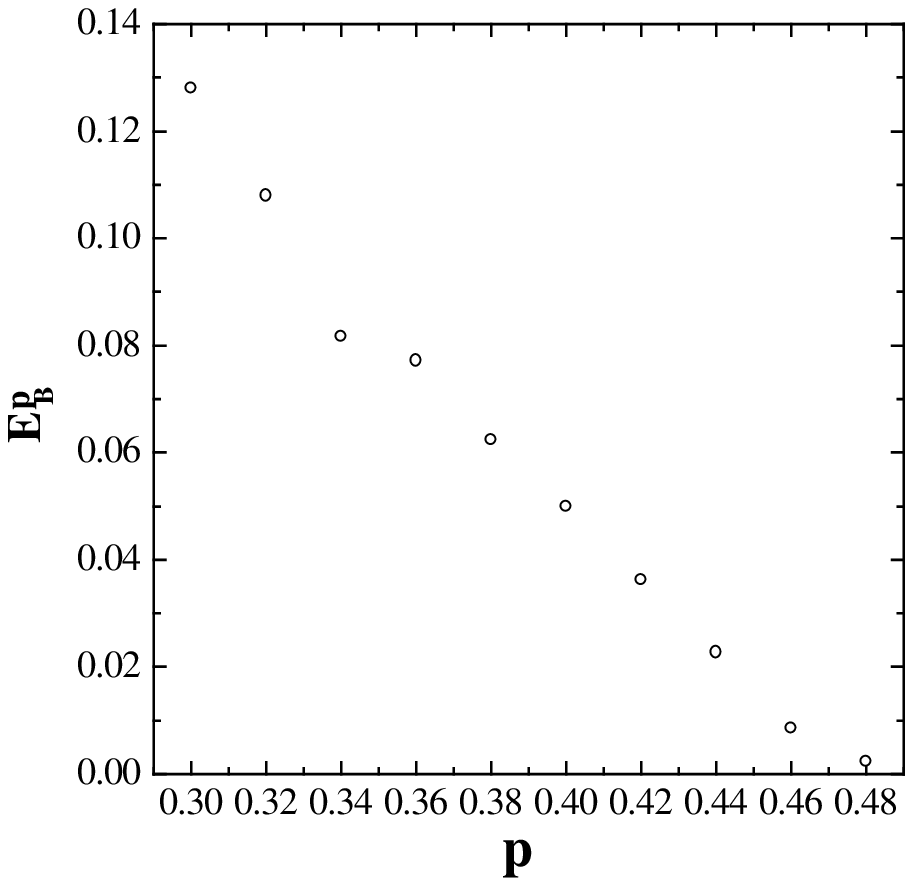}
\end{figure}

\begin{figure}
\psfig{file=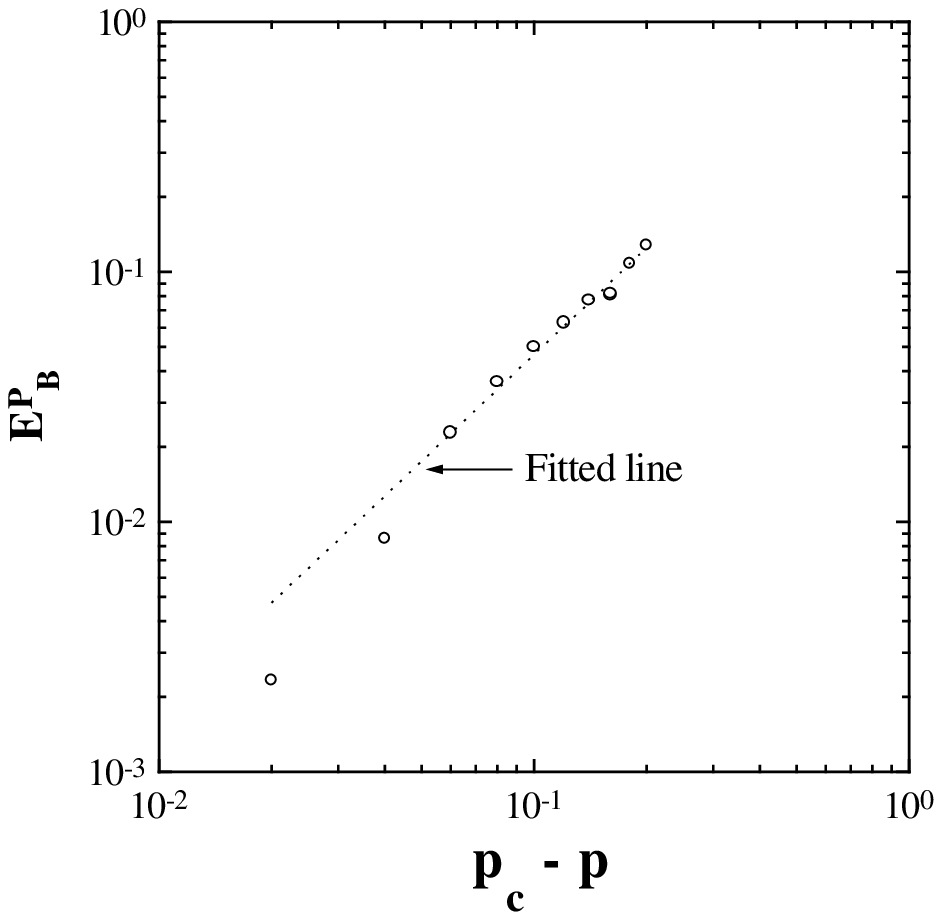}
\end{figure}

\end{document}